\documentclass[sigconf]{acmart}
\usepackage{rotating}
\usepackage{subfig}
\usepackage{color}
\usepackage{soul}
\usepackage{multirow}

\AtBeginDocument{%
  \providecommand\BibTeX{{%
    \normalfont B\kern-0.5em{\scshape i\kern-0.25em b}\kern-0.8em\TeX}}}

\copyrightyear{2024}
\acmYear{2024}
\setcopyright{rightsretained}
\acmConference[CHI '24]{Proceedings of the CHI Conference on Human Factors in Computing Systems}{May 11--16, 2024}{Honolulu, HI, USA}
\acmBooktitle{Proceedings of the CHI Conference on Human Factors in Computing Systems (CHI '24), May 11--16, 2024, Honolulu, HI, USA}
\acmDOI{10.1145/3613904.3642490}
\acmISBN{979-8-4007-0330-0/24/05}

\begin{document}

\title{Viblio: Introducing Credibility Signals and Citations to Video-Sharing Platforms}

\author{Emelia May Hughes}
\email{ehughes8@nd.edu}
\affiliation{%
  \institution{University of Notre Dame}
  \city{South Bend}
  \state{Indiana}
  \country{USA}
}

\author{Renee Wang}
\email{renee17@cs.washington.edu}
\affiliation{%
  \institution{University of Washington}
  \city{Seattle}
  \state{Washington}
  \country{USA}
  }
  
\author{Prerna Juneja}
\email{prerna79@uw.edu}
\affiliation{%
  \institution{University of Washington}
  \city{Seattle}
  \state{Washington}
  \country{USA}
  }
  
\author{Tony Li}
\email{toli@ucsd.edu}
\affiliation{%
  \institution{University of California San Diego}
  \city{San Diego}
  \state{California}
  \country{USA}
  }

\author{Tanu Mitra}
\email{tmitra@uw.edu}
\affiliation{%
  \institution{University of Washington}
  \city{Seattle}
  \state{Washington}
  \country{USA}
  }

\author{Amy Zhang}
\email{axz@cs.uw.edu}
\affiliation{%
  \institution{University of Washington}
  \city{Seattle}
  \state{Washington}
  \country{USA}
  }

\begin{abstract}
As more users turn to video-sharing platforms like YouTube as an information source, they may consume misinformation despite their best efforts. In this work, we investigate ways that users can better assess the credibility of videos by first exploring how users currently determine credibility using existing signals on platforms and then by introducing and evaluating new credibility-based signals. We conducted 12 contextual inquiry interviews with YouTube users, determining that participants used a combination of existing signals, such as the channel name, the production quality, and prior knowledge, to evaluate credibility, yet sometimes stumbled in their efforts to do so. We then developed Viblio, a prototype system that enables YouTube users to view and add citations and related information while watching a video based on our participants' needs. From an evaluation with 12 people, all participants found Viblio to be intuitive and useful in the process of evaluating a video’s credibility and could see themselves using Viblio in the future. 
\end{abstract}

\ccsdesc[500]{Human-centered computing~Social tagging systems}
\ccsdesc[300]{Human-centered computing~Reputation systems}
\ccsdesc[500]{Human-centered computing~Social networking sites}
\ccsdesc[300]{Human-centered computing~Ethnographic studies}
\ccsdesc[300]{Human-centered computing~Social content sharing}
\ccsdesc[300]{Human-centered computing~Collaborative filtering}
\ccsdesc[300]{Human-centered computing~Social media}
\ccsdesc[300]{Human-centered computing~Social tagging}

\keywords{Credibility Signals, Citations, YouTube, Misinformation, Social Platforms, Contextual Inquiry, Semi-Structured Interview}



\maketitle

\section{Introduction}\label{intro}
Social media has simultaneously lowered the barrier to publishing content and aided the spread of misinformation online. YouTube in particular has created both a space for individuals to share and consume a variety of informative and entertaining media as well as a platform for bad actors to disseminate misinformation \cite{axelrod_youtube_2022, tokojima_machado_it-which-must-not-be-named_2022}. Traditional search engine results tend to promote mainstream sources that have the authority and content favored by search engine algorithms. On YouTube, however, search results are affected by user watch history \cite{hussein_measuring_2020}, which creates a space for non-authoritative content creators and people spreading misinformation and conspiracy theories. 

Researchers have investigated a variety of countermeasures to misinformation on social media platforms, given the widespread concern \cite{argentino_qanon_2021, bengali_how_2019, coleman_hundreds_2020, reed_hate_2018}.  Prior initiatives to combat misinformation have taken many forms, including detecting false or misleading information using machine learning algorithms \cite{castillo_information_2011, potthast_clickbait_2016, shu_fake_2017}, crowd-sourcing tools \cite{allen_scaling_2020, bhuiyan_investigating_2020, epstein_will_2020, kim_leveraging_2018, pennycook_fighting_2019}, and providing fact-checked information related to circulated news claims \cite{moran_deciding_2018, kriplean_integrating_2014, pennycook_implied_2020, yaqub_effects_2020}.

\begin{figure*}
\centering
\includegraphics[width=6in]{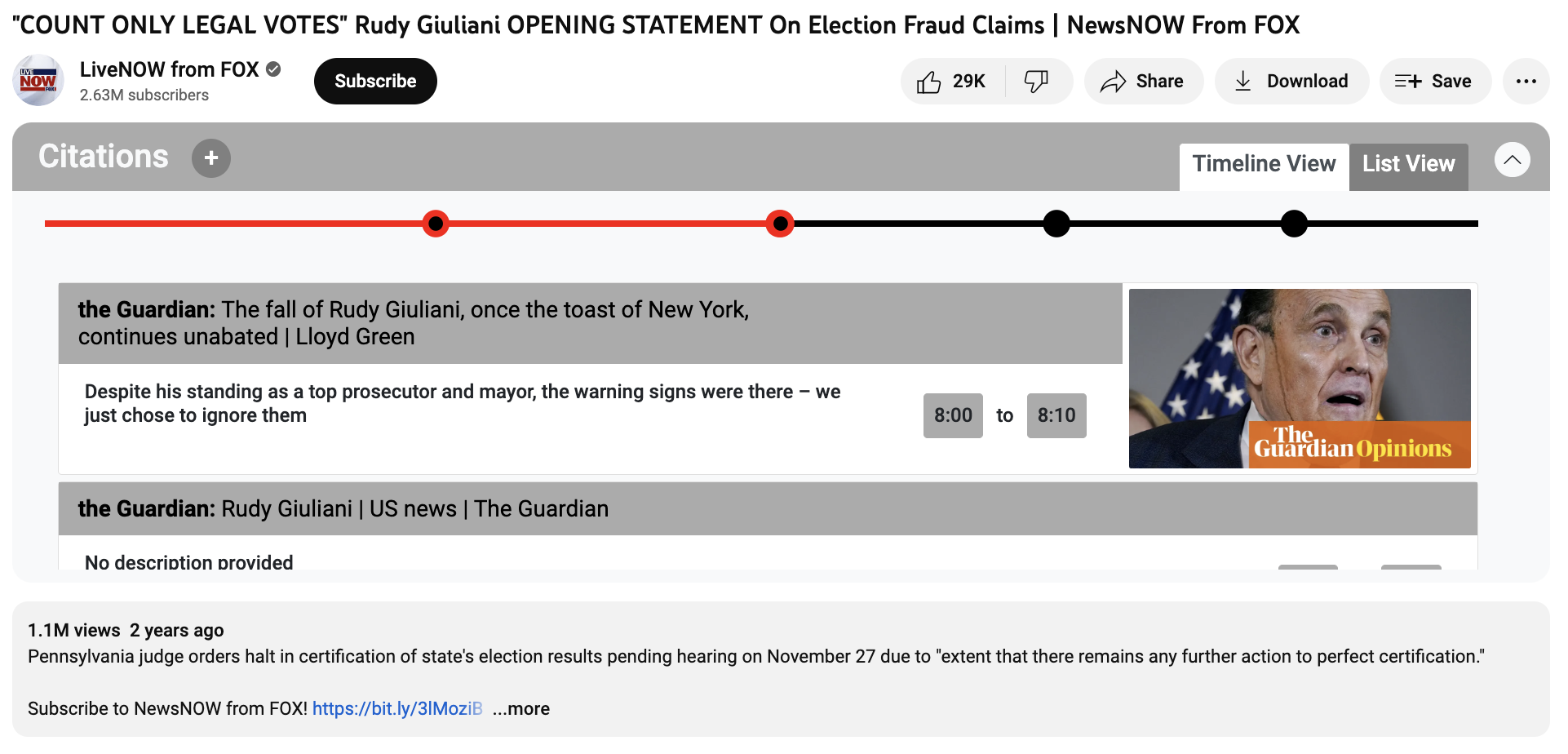}
\caption{Citations added by study participants on a controversial video, shown in Viblio's timeline-view.}
\label{timeline-view}
\end{figure*}

These approaches mainly focus on stopping the spread of misinformation but do not confront the underlying factors that prevent users from determining the credibility of content, like YouTube videos, in the first place. More recently, researchers and platforms have examined additional methods that give users greater context about the content they see, so that they can spot instances of misinformation for themselves~\cite{Zhang2018,10.1145/2145204.2145274,facebook,twitter,10.1145/3610061}. These approaches are complementary to take-down forms of moderation as they give users the flexibility to apply their own credibility standards and trust measures to evaluate content. 

Prior research has highlighted a wide range of possible indicators for evaluating content credibility such as visual appearance, tone, and representative citations \cite{Zhang2018,10.1145/2145204.2145274}. In particular, Wikipedia and Wikidata have emerged as key players in the space of credibility signals, as many platforms have turned towards signals from or display links to Wikipedia. These indicators help users gain greater context about a topic by engaging in lateral reading practices, where readers cross-reference external materials while reading the original source. Lateral readers have been shown to gain a better sense as to whether to trust the facts and analysis presented to them \cite{Caulfield2017,wineburg2017lateral}.
 
However, it is still unclear which types of credibility indicators are most useful to online video-based information consumers and how they should be presented. In this work, we explore how credibility is conveyed on video-sharing platforms, due to the lack of standard credibility factors or citation displays in such spaces. For example, a search query made through a web search engine will result in  Knowledge Panels and Wikipedia snippets; yet, these signals are missing when the same query is made on a video-sharing platform like YouTube. This lack of general credibility signals is concerning, considering that video-sharing platforms, like YouTube and TikTok, have been shown to be used as informational sources for many users \cite{pewresearch,smith_many_2018,ManyAmer34:online}. In particular, young people are increasingly turning to social platforms for information, as they seek a socially-informed awareness of the value of the information they encounter \cite{10.1145/3544548.3581328, nytimes}.

Thus, we explore how users currently determine the credibility of videos they watch on video sharing platforms and how this process can be improved. We first conducted a contextual inquiry study to understand how users interact with existing credibility signals on YouTube. Through 12 interviews with YouTube users, we explored what factors help them select and trust a video, what signals (if any) might be confusing or misleading to them, and where they could imagine any additional credibility signals introduced on YouTube. 

Across all interviews, all participants referenced source familiarity, channel name, and production quality in selecting a YouTube video from a search query. The most common (mentioned by 8/12 participants) signals on the playback page referenced for trusting a video were the video description and the video comments section, echoing prior work on the importance of social signals towards credibility \cite{10.1145/3331184.3331285, 10.1145/3544548.3581328}. We noticed that among the signals that participants mentioned, most were imperfect proxies that could be misinterpreted. One participant even interpreted YouTube's information panel, which specifically populates on topics prone to misinformation with the goal of countering it~\cite{GoogleHelp}, as an endorsement for the content within the video. Another imperfect signal users referenced is the production quality of the video. While production quality does not necessarily mean that the video is presenting correct information, some users interpret that as such. The results of our contextual inquiry emphasized the need for a tool that integrates more information-rich credibility signals and promotes accessible lateral reading in a socially-informed way.

Building on the findings from our contextual inquiry, previous work, and existing credibility-based systems, we built Viblio, a prototype system that allows users to add and view citations on YouTube videos. The results of our initial study emphasized that existing signals are simply weak proxies for credibility, so we set out to build a system that intentionally centers credibility. Viblio allows any YouTube user, not just the video creator, to insert citations along the video timeline and intentionally interact with credibility-related information. These citations are displayed dynamically during video playback, enhancing the viewers' ability to access additional information. The extension is conveniently located below the video and alongside channel information, facilitating the crowd-sourcing of citations for a wide range of videos.

To evaluate our system's efficacy as a credibility signal on YouTube, we tested Viblio in an extended user study. Our goal with this study was to evaluate the effect of Viblio on participants, the integration (or lack thereof) of Viblio with existing credibility signals, and the topics where Viblio could be of use to YouTube users. Over the period of about 15 days, 12 participants used Viblio and reported on the extension as well as its impact on their credibility ratings on videos.
Participants were given 2--3 videos to watch and interact with each day. Participants rated the credibility of each video on a scale of 1--5 both before exploring citations and the video, and then again after. The study concluded with interviews, where participants reflected upon their experience with and the usability of Viblio. Overall, participants found Viblio intuitive to use and helpful for developing a greater understanding and context for the videos they watched. Notably, participants saw great potential for use in educational contexts as well as across different social platforms. Generally, Viblio was found to be useful in controversial and political topics; however, concern was raised about possible misuse by bad actors.  

The main contributions of this work encompass:
\begin{itemize}
    \item An interview rooted in contextual-inquiry methods, which uncovered gaps in users' ability to assess credibility on YouTube due to the absence of intentional credibility signals.
    \item Viblio, a prototype system built as an extension for YouTube that allows users to add, view, and interact with external sources related to the video content.
    \item A user study with 12 participants that validated the overall usability and effectiveness of Viblio, as well as introduced new possible applications for citation-based tags on shared videos. 
\end{itemize}

\section{Background and Related Work}\label{background}
\subsection{Study context: YouTube Platform}
YouTube is the leading video-sharing platform with 2.6 billion users worldwide \cite{YouTubeR25:online,HowGoogl14:online}. While YouTube serves as an important source of information, with one in four US adults turning to the platform for their news \cite{ManyAmer34:online}, it is also a fertile ground for misinformation and conspiracy theories about health, vaccines, politics, climate change, social issues, and more~\cite{hussein_measuring_2020,article_climate,juneja2023assessing,10.1145/3555143}. To compound the issue, the platform's recommendation algorithms have been criticized for driving people down ``rabbit holes'' of misinformative content \cite{hussein_measuring_2020, hosseinmardi2023causally}.  In the past, YouTube has tried to tackle the problem by removing misinformative videos from the platform, pushing mainstream channels as top search results, suppressing the recommendations of problematic information, and demonetizing misinformation-spreading channels \cite{article9,YouTubeu55:online,Demoneti53:online}. YouTube also introduced informational interventions, such as the inclusion of Wikipedia links within videos promoting conspiracy theories \cite{YouTubea93:online} and the integration of fact-checks from independent organizations into search results \cite{HowGoogl52:online}. Some of these interventions have demonstrated efficacy, reducing traffic to anti-vaccine videos on the platform \cite{kim2020effects}. However, despite these efforts, YouTube continues to grapple with misinformation. In this research, we aim to address this issue by exploring methods to increase user awareness of underlying inaccuracies in videos. Our approach focuses on combating misinformation in the case that it does reach viewers by allowing them to explore intentional credibility-related signals and practice lateral reading techniques.

\subsection{Social media signals affecting credibility perceptions}

In the digital era, the internet has brought forth an abundance of information sources, offering people an extensive array of options for finding information. Faced with this deluge of information, people often exert little cognitive effort in deeply reflecting on the validity of online information and instead may rely on heuristics to make assessments about the content \cite{chaiken2014heuristic,walther2012communication,petty1986elaboration}. Such heuristics are often triggered by online signals or cues such as structural characteristics (e.g., length of content), communicator characteristics (e.g., likability, appearance), or audience characteristics (e.g., audience's reactions) \cite{chaiken2014heuristic}. Scholars have posited that there are various cues embedded in the affordances (modality, agency, interactivity, navigability) provided by online platforms that can activate cognitive heuristics shaping users' perception about the credibility of online content \cite{sundar2008cognitive}. When a group of users collectively like or share a news article on social media, these actions signal collective trust in the information's credibility \cite{hong2006influence,flanagin2008digital}. Additionally, agency affordance can also activate the authority heuristic when users perceive the source of information as authoritative. This occurs because individuals are inclined to place more trust in content originating from experts, reputable organizations, or authoritative accounts \cite{briggs2002trust,lin2016social}.  

While online platforms present numerous signals across all four aforementioned affordances, not all of these signals lead to sound judgments. Research has demonstrated that people may form inaccurate judgments based on easily manipulable signals, such as the number of sources cited in an article \cite{sundar1998effect} or visual aesthetics \cite{metzger2010social}, or during the sensemaking process \cite{10.1145/3134696, 10.1145/3025453.3026012, 10.1145/3290605.3300563}. Therefore, it is important to identify and understand the role of diverse signals present on online platforms that influence users' perceptions of online credibility. On top of the signals present on online platforms, recent work using Community Notes (formerly Birdwatch) has found that contextual features, such as the partisanship of users, can affect a user's judgment of whether other users' tweets are misleading \cite{10.1145/3491102.3502040}. While prior research has explored credibility signals on platforms such as X (formerly Twitter) \cite{edgerly2019blue,mohd2014user,lin2016social,mohd2014user} and Facebook \cite{ali2022fake,borah2018importance}, as well as on news websites \cite{sundar2007news,pjesivac2018social}, and the broader web ecosystem (see \cite{choi2015web} for a review), there exists a significant gap in research concerning video-based platforms like YouTube. While previous work has studied the features of video content that affect users' engagement with videos \cite{munaro2021engage}, we still lack an understanding of how users evaluate the credibility of content on this type of platform. Our exploratory study aims to bridge this gap by identifying the diverse signals on YouTube that users rely on to evaluate and trust video content, while also highlighting the challenges users face with existing credibility signals.

\subsection{Interventions to combat online misinformation}
To combat online misinformation, both scholars and online platforms have employed a wide array of strategies and approaches \cite{aghajari2023reviewing}. These include establishing policies against misinformation for platform governance \cite{Alongter67:online,YouTubeu55:online,Facebook65:online}, reducing the visibility of misleading content by suppressing and down-ranking it from searches and recommendations \cite{Facebook18:online,epstein2020will,Facebook81:online}, as well as identifying and removing misinformative content \cite{Pinteres3:online,Facebook10:online,Facebook99:online}. Platforms have also embraced measures like de-platforming and demonetization to limit the influence of repeat spreaders of misleading content \cite{Removing47:online,TikTokis90:online,Facebook75:online}. Concurrently, efforts to raise public awareness about misinformation have been amplified through targeted media literacy programs \cite{sirlin2021digital,jones2021does,buchanan2020people}. Furthermore, scholars have also explored several design interventions to aid individuals in assessing the credibility of online content \cite{bhuiyan2021nudgecred,TwitterN50:online,kirchner2020countering,spezzano2021s,schwarz2011augmenting,dias2020emphasizing,10.1145/3375189}. They include interventions targeting online content, including message-based approaches conveying the detrimental consequences of misinformation \cite{chen2015deterring}, incorporation of warning messages \cite{TwitterN50:online,kirchner2020countering},  inoculation messages \cite{innoculation}, and debunking of false information through fact-checks and corrective messages \cite{bush2019implications,garrett2013promise,boukes2023fighting,carey2022ephemeral}. Recognizing that the source of information plays a pivotal role in shaping individuals' credibility assessments, several design interventions seek to redirect attention towards source information and the authoritativeness of the source itself \cite{bhuiyan2021nudgecred,spezzano2021s,schwarz2011augmenting,dias2020emphasizing}. Furthermore, scholars have explored the potential of crowdsourced judgments in evaluating the credibility of online information \cite{bhuiyan2021nudgecred,huang2013credibleweb,borwankar2022democratization}. Crowd-based interventions have also been successfully adopted by online platforms. For instance, X launched the Community Notes initiative that allows users to identify and annotate tweets that they believe contain misinformation or false content \cite{Introduc53:online}. 

In our work, we built a prototype system called Viblio which leverages crowdsourcing to incorporate credibility citations directly into YouTube videos, providing users with additional contextual information while they are viewing video content. Along with results from our exploratory study, we took inspiration from Community Notes and Wikipedia to inform the design of Viblio. Community Notes provided an example of how crowdsourcing could directly influence both viewer's trust in a source as well as their active engagement in incorporating credibility awareness into their usual viewing habits. Wikipedia showed us how citations can impact trust \cite{wikiarticle} and allowed us to envision how citations could be embedded within a video as well as what type of videos could benefit from citations.

\section{Exploratory Study Overview}\label{exploratory-study}
Prior to designing our system, we conducted a contextual inquiry to understand how users currently interact with YouTube and its existing credibility signals. We also wanted to explore the implementation of citations as a promising method for increasing information context on YouTube, especially looking at the success of Community Notes and Wikipedia in this area. For this study, we developed 3 research questions (RQs):
\begin{enumerate}
    \item What existing credibility signals on YouTube do viewers use to determine what videos to watch as well as trust?
    \item What problems do users have with existing credibility signals on YouTube?
    \item What are potentially promising signals for credibility that are not currently present on YouTube?
\end{enumerate}

\subsection{Methods}
Participants were primarily recruited through word-of-mouth and X. Participants were asked about basic demographic information like age range, gender, and education level, and were then screened for comfort with screen sharing and being recorded in online meetings (Table \ref{explorative-participants}). Generally, participants were selected by sign-up order and to diversify the participant group as much as possible. Of the 12 participants (Table 1), the majority were in the 18-22 age range and currently in an undergraduate program (7 out of 12). Overall, the participants skewed female (8 out of 12) and were under the age of 30 (10 out of 12). Recruitment materials presented the study as one focused on misinformation and credibility on YouTube. Each participant was compensated \$15 for their time.

\begin{table}[b]
\caption{Participant Demographics in the Contextual Inquiry}
\begin{tabular}{c|c|c|c}
\hline\textbf{Participant} & \textbf{Age Range} & \textbf{Gender} & \textbf{Education} \\ \hline
1                    & 18-22              & Female          & In college         \\
2                    & 18-22              & Male            & In college         \\
3                    & 23-30              & Female          & In graduate school \\
4                    & 18-22              & Female          & In college         \\
5                    & 18-22              & Male            & In college         \\
6                    & 18-22              & Male            & In college         \\
7                    & 18-22              & Male            & In college         \\
8                    & 18-22              & Female          & In college         \\
9                    & 45-60              & Female          & Master's or above  \\
10                   & 31-45              & Female          & Bachelor's         \\
11                   & 23-30              & Female          & Master's or above  \\
12                   & 23-30              & Female          & Master's or above \label{explorative-participants} \\ \hline 
\end{tabular}
\end{table}

With IRB approval from our institution (STUDY00013111), 
interviews were conducted virtually over Zoom across a 3--week period. Each interview lasted approximately 60 minutes and was recorded and transcribed using Zoom’s recording software. Before conducting each interview, the participant was asked for consent to record the session and told that they could stop the interview at any time if needed. All searches were done in an incognito browser window on the participant’s computer to prevent the participant’s YouTube or Google recommendations from influencing the study. This study was conducted prior to YouTube’s removal of dislike counts on videos.
The format of the interviews borrows from contextual inquiry methods and a semi-structured interview format. Participants were given 4 tasks to complete, in order: (1) video selection using results page signals, (2) comparing YouTube and Google results, (3) video credibility using playback page signals, and (4) citation generation while watching videos. Once each of the 4 tasks were completed, we asked participants open-ended questions about their process and choices. At the end of each interview, we asked participants general questions related to their YouTube usage and their knowledge of citations and credibility. 
The activities participants were asked to complete are as follows:
\begin{itemize}
    \item Video selection using results page signals: Participants were given four phrases to search on YouTube. For each search term, participants were asked to think out loud while they selected a video to watch on that topic, explaining why they would choose one video over another. Chosen for their variety in terms of controversy level and need for credible information, the search terms were given in the order: ``Europe travel guide'', ``should I go vegan?'', ``climate change 2050'', and ``different COVID vaccines''.
    \item Comparing YouTube and Google results: Participants were asked to compare and contrast the search results for ``different COVID vaccines'' on YouTube and Google, focusing on the user interface, any signals that stood out, and what they liked/disliked on both platforms. Finally, they were asked about their preference for one platform over another as a source of information and entertainment and the reasoning behind such a preference if one existed.
    \item Video credibility using playback page signals: Participants opened the first two YouTube results for ``different COVID vaccines''. Without watching the videos, we asked participants to identify why they believed one video may be more credible or trustworthy than the other. This exercise let us determine what signals exterior to the video participants used to evaluate credibility.
    \item Citation generation while watching videos: Participants were asked to watch 2 pre-selected videos on YouTube. There were three possible video combinations given to participants:
        \begin{itemize}
            \item A video supporting and a video disproving the moon landing conspiracy theory.
            \item An anti-vaccination video and a video disproving claims about links between vaccines and autism from the Mayo Clinic.
            \item Election fraud in the 2020 election from CBS and FOX.
        \end{itemize}
    \item Participants were asked to write down the timestamps where they thought a citation or further information would be necessary. For each point, they were asked to find a source online that they believed could be cited. Participants were free to format the citation however they chose. 
\end{itemize}

The three interviewers used open coding methods to perform a preliminary round of independent qualitative coding using the interview transcripts. After discussion, a consolidated code--book was established and used by the primary interviewer to code the transcripts. Based on the code--book, emerging topics were identified by thematic analysis. Codes included but were not limited to result order on the search results page, video length, video channel verification, video thumbnail and preview, date posted, title, production quality, description box, comments, suggested videos, production quality, and resemblance to clickbait (sensationalized title and/or thumbnail).

\subsection{Results}

\begin{table*}[t]
\small
\caption{All signals mentioned by four or more participants when describing whether to trust a video or not.}
\begin{tabular}{l|l|l|l}

\hline
\label{explorative-results}\textbf{Credibility Signal} & \textbf{Description}                                                                                                          & \textbf{Count} & \textbf{Representative Quote}                                                                                                                                                                                                                                                                                                                                                                    \\ \hline
Source familiarity          & \begin{tabular}[c]{@{}l@{}}Talking about having heard \\ of the channel before.\end{tabular}                                  & 12/12          & \begin{tabular}[c]{@{}l@{}}``...well this is something that I heard of. I know about it a little\\ bit, so I would probably click on that'' (P6).\end{tabular}                                                                                                                                                                                                                                     \\ \hline
Channel name                & \begin{tabular}[c]{@{}l@{}}Looking at the channel that \\ posted the video.\end{tabular}                                      & 12/12          & \begin{tabular}[c]{@{}l@{}}``I would look through these, okay, and the BBC and the \\ Today Show are going to be a much vaguer explanation...So \\ the AMA, okay, so that is the American Medical Association, \\ that might be of interest as well...Okay you see Davis Health \\ I know that’s the University of California, Davis, so that might \\ be of interest too'' (P9).\end{tabular}     \\ \hline
Production quality          & \begin{tabular}[c]{@{}l@{}}Video production quality, \\ as perceived by the viewer.\end{tabular}                              & 12/12          & \begin{tabular}[c]{@{}l@{}}``There’s not a lot of production put into it, which is starting to \\ make me doubt the content quality'' (P2).\end{tabular}                                                                                                                                                                                                                                           \\ \hline
Video description           & \begin{tabular}[c]{@{}l@{}}The description\\ accompanying the video\end{tabular}                                              & 8/12           & \begin{tabular}[c]{@{}l@{}}``Usually I always try and start by looking at the description \\ because, to me, a person who wants the video to be credible \\ would put a good detailed description here. . . they’ll have a \\ good enough description to get people interested in their \\ video and get some information if they don’t have time to \\ watch the entire thing'' P5).\end{tabular} \\ \hline
Interest                    & \begin{tabular}[c]{@{}l@{}}Thinking a video looks \\ interesting or boring.\end{tabular}                                      & 7/12           & \begin{tabular}[c]{@{}l@{}}``Chances are that I’m on YouTube because I am looking \\ for something that’s entertaining'' (P12).\end{tabular}                                                                                                                                                                                                                                                       \\ \hline
Thumbnail                   & \begin{tabular}[c]{@{}l@{}}Using the video thumbnail \\ on the results page.\end{tabular}                                     & 6/12           & \begin{tabular}[c]{@{}l@{}}``I don’t like how this thumbnail looks. It looks like it’s a little\\ outdated, I don’t know, I just don’t like the look of it'' (P4).\end{tabular}                                                                                                                                                                                                                    \\ \hline
Date posted                 & Date posted on YouTube.                                                                                                       & 5/12           & \begin{tabular}[c]{@{}l@{}}``I mean, if it’s really old, If it’s maybe three or four years \\ old, it may not be up to date with what’s really going on'' (P6).\end{tabular}                                                                                                                                                                                                                       \\ \hline
Title                       & \begin{tabular}[c]{@{}l@{}}Title of the video as \\ displayed on the results page.\end{tabular}                               & 5/12           & \begin{tabular}[c]{@{}l@{}}``I am somebody who just doesn’t really prefer any sort of \\ like click bait or very catchy titles, I think that kind of thing \\ is very misleading'' (P7).\end{tabular}                                                                                                                                                                                              \\ \hline
Video preview               & \begin{tabular}[c]{@{}l@{}}Video preview that appears \\ when the user hovers over \\ the video thumbnail.\end{tabular}       & 5/12           & \begin{tabular}[c]{@{}l@{}}``I mean this feature, a very mini preview of it {[}the video{]}, \\ I think that is also kind of important. So for example if I’m \\ looking at this, this seems more like a monologue kind \\ of thing'' (P6).\end{tabular}                                                                                                                                           \\ \hline
Search result order         & \begin{tabular}[c]{@{}l@{}}Basing decision on results \\ ranking algorithm.\end{tabular}                                      & 5/12           & \begin{tabular}[c]{@{}l@{}}``I would assume that YouTube would be putting, like, let’s \\ say the more official sources or they will be promoting the \\ videos from the official sources, up front'' (P12).\end{tabular}                                                                                                                                                                          \\ \hline
Comments section            & The comments on the video                                                                                                     & 5/12           & \begin{tabular}[c]{@{}l@{}}So if I were to see troll comments. . . I would just go to a \\ different video because I don’t want to waste my time \\ looking at a video that might not be credible'' (P12).\end{tabular}                                                                                                                                                                           \\ \hline
Number of views             & \begin{tabular}[c]{@{}l@{}}Number of views on the \\ video.\end{tabular}                                                      & 5/12           & \begin{tabular}[c]{@{}l@{}}``So if you have a high number of views It might indicate that\\ hey, a lot of people have watched it, and it might be relevant \\ and it might have some more trustworthy content'' (P6).\end{tabular}                                                                                                                                                                 \\ \hline
Video length                & \begin{tabular}[c]{@{}l@{}}Video duration as shown \\ on the results page.\end{tabular}                                       & 4/12           & \begin{tabular}[c]{@{}l@{}}``I’m just kind of looking for a video that’s not too long but \\ also has good information that I could use to learn about \\ climate change'' (P8).\end{tabular}                                                                                                                                                                                                      \\ \hline
Likes and dislikes          & \begin{tabular}[c]{@{}l@{}}The number of likes or \\ dislikes that the video has, \\ and the ratio between them.\end{tabular} & 4/12           & \begin{tabular}[c]{@{}l@{}}For all you know these dislikes could come from many \\ factors and that doesn’t mean that the video is not \\ credible at all'' (P12).\end{tabular}                                                                                                                                                                                                                   \\ \hline
Subscriber count            & \begin{tabular}[c]{@{}l@{}}The number of subscribers \\ that the video creator has.\end{tabular}                              & 4/12           & \begin{tabular}[c]{@{}l@{}}``I think the low subscriber rate and the lack of production \\ of the video has made me a little shifty on this'' (P2).\end{tabular}                                                                                                                                                                                                                                   \\ \hline
\end{tabular}
\end{table*}

In this study, we found several signals in which participants commonly relied upon (Table \ref{explorative-results}). The most popular included familiarity with the YouTube channel, production quality, and the video description. It was also revealed, however, that the participants were skeptical of the effectiveness and reliability of certain signals. Some signals, such as YouTube’s verification system and popup banners, were misinterpreted by some participants. Participants also suggested potential improvements and new approaches to credibility on YouTube, including knowledge panels on search results pages and introducing citations. 

\subsubsection{Issues With Existing Credibility Signals}
In the current approach to evaluating credibility on YouTube, users commonly rely on several signals. While all signals mentioned by four or more participants are shown in Table \ref{explorative-results}, we elaborate on signals that provide better insight into participants' process of determining credibility below. These include:

\paragraph{Search results ranking} Some participants (5/12) mentioned that the order in which videos appear, as determined by the YouTube ranking algorithm, affected their video selection process. Most believed that YouTube ranks videos based on relevance, but some, like P12, believed that YouTube promotes videos with certain sources and ranks them higher in the search results: ``\textit{I would assume that YouTube would be putting, like, let’s say the more official sources or they will be promoting the videos from the official sources, up front}''. YouTube’s ranking algorithm is complex; our study was done in incognito mode, but watch history affects how YouTube ranks videos for the many users who do not use incognito mode. Thus, users might be misinterpreting YouTube’s ranking algorithm, which is concerning as it has been shown that watching videos with misinformation can lead to more recommendations with misinformation \cite{hussein_measuring_2020, boyd_data_2018}. One participant, P4, compared YouTube’s ranking algorithm to Google’s search page algorithm. They felt that Google’s algorithm was more purposeful in ranking relevant results first, while YouTube’s algorithm was guessing at what they wanted to see. ``\textit{With YouTube, I don’t know exactly how the algorithm works...it’s not like it shows the verified ones most or it’s not like it shows the most viewed ones most. Maybe it’s just guessing what I would like. I think in terms of the ordering of things I agree with the Google search one more, just because it kind of feels like things are ranked based on how technical things are}''.

\paragraph{Video metadata} Video length, posting date, and view count influenced video selection, with view count often associated with credibility. Subscriber count and channel verification played a role in determining credibility. Participants also used user engagement metrics such as view count (3/12), like/dislike counts (4/12), and comments (8/12) to decide whether or not to trust a video. Among the participants who cited view count as a reason to trust the video, all of them thought that more views implied a more credible creator. Participant P11 also associated view count with subscriber count and thought that together, they could imply that the source is credible: ``\textit{This is something related to a very renowned show and has a lot many subscribers, 10.8 million subscribers and around 1 million views...so a good source}''. Participants generally agreed that the number of dislikes on a video did not lower video credibility for them: ``\textit{For all you know these dislikes could come from many factors and that doesn't mean that the video is not credible at all}'' (P12). Participants sometimes also referenced the comments as credibility signals, where seeing ``troll'' comments might discourage viewers from trusting the video: ``\textit{So if I were to see troll comments...I would just go to a different video because I don’t want to waste my time looking at a video that might not be credible}'' (P12). Troll comments also have an effect on how participants trust other signals on the playback page, like the like/dislike count. As P2 said, ``\textit{Right off the bat, seeing that there’s a huge dislike to like ratio is something of concern. And then I would go down to the comments, then honestly reading these comments, I’m starting to think that the dislikes are more from people that don’t understand the content rather than being like this is not factual.}'' 

\paragraph{Channel verification} Across different exercises in our study, participants noted that there was a lack of clarity on how YouTube channel verification works. Participants, like P8, expressed general confusion over the process: ``\textit{Yeah, I mean I don’t really know the specifics [of the verified channel check] but I don’t really think that it is related to the validity of their content that they provide, but I think that it definitely increases their credibility in a sense that it’s not like fake or it’s not just like a scam or something}''. Beyond the confusion, some participants also had a misconception about how verification works and interpreted verification as a definitive credibility signal. For example, P9 referenced channel verification as a reason for trusting a video, saying that ``\textit{this channel is verified so I know that it’s not, like, providing false information}''. The YouTube channel verification process does not involve content verification at all \cite{noauthor_verification_nodate}, so the verification check can be misleading for YouTube users who are looking for credibility signals for video content.

\begin{figure*}
\centering
\includegraphics[width=5in]{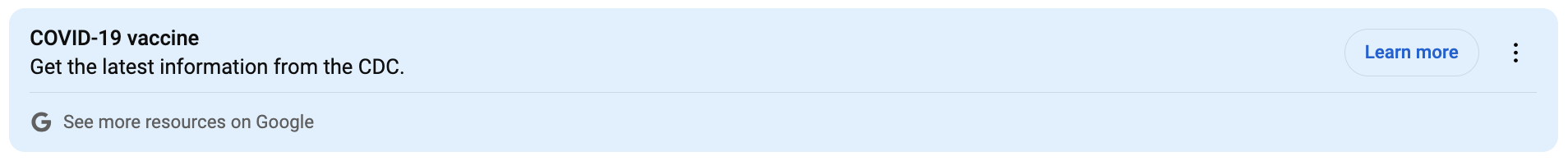}
\caption{YouTube's information panel on videos related to COVID-19}\label{YT-info-panel}
\end{figure*}

\paragraph{Production quality} Production quality was also brought up by all participants, where some participants, like P2, related video quality to content quality: ``\textit{There’s not a lot of production put into it, which is starting to make me doubt the content quality}''. Having high production quality could act as a credibility signal even when the channel source was unfamiliar. For instance, P3 did not rule out a video even though they did not know of the source, saying that ``\textit{I don’t know who that is, but it looks pretty well designed and well funded. I don’t fully trust it because I don’t know who they are, but they look like they have good production quality [in the video], so we’ll see}''. We saw that production quality can compensate for a lack of other credibility signals, such as channel familiarity and verification.

\paragraph{Misinformation--focused information panels} Participants identified multiple problems with YouTube’s information panels, which they confused with ads. YouTube's information panels appear on search results and video topics ``prone to misinformation, such as the moon landing''~\cite{GoogleHelp}. One example is the COVID panel that accompanies videos on COVID-19 that links to the CDC website (Fig. \ref{YT-info-panel}). One participant, P5, noted that ``\textit{One of the things that I want to say about YouTube that I don’t like is that a lot of times they use these pop ups to also display ads. I usually tend to ignore these because I just believe they’re all ads, but like in this is clearly not an ad, this is like something that I can use to go to a CDC website}''. Thus, the design of YouTube’s informational panel can unintentionally dissuade viewers from looking at it. Another issue with the panel was that it was being misinterpreted. One user interpreted it as an endorsement by the CDC. P9 saw the panel under a video on COVID-19 vaccines and said, ``\textit{...get the latest info from the CDC [referring to the popup]. So it’s telling me where it’s getting its information from, which coming from the CDC would make [the video] a little more trustworthy to me}''. In this instance, YouTube’s credibility indicator was not interpreted the way it was intended---indeed, it had the opposite intended effect for at least one of our participants---suggesting that we should be cautious when introducing new credibility indicators so that they are not misinterpreted by users in a similar manner.

\subsubsection{New Approaches to Credibility}
As for new approaches to credibility, participants suggested numerous potential improvements.

\paragraph{Search result knowledge panels} Potential improvements to YouTube’s search results page were identified when participants were asked to make the same search query on YouTube and Google. P4 noted that Google supplied a lot of detailed quantitative information on the search query but thought that the same information would be too much for YouTube. Instead, P4 thought that YouTube could benefit from a knowledge panel with general information: ``\textit{I think all this chart stuff [on Google] is a little bit much for the YouTube platform specifically but I do like [the knowledge panel] $\ldots$ [it would be] helpful to have on the side, just so that it’s accessible but also people $\ldots$ on YouTube might like learn something on the side that they weren't expecting to}''. While adding knowledge panels to search result pages could be beneficial, there is a risk of confusion similar to YouTube's current information panels. A distinction from channel-provided content would need to be made, and the scope should spread beyond that of misinformation-focused information panels.

\paragraph{Citations}\label{citationquote} Based on the success of Wikipedia and Community Notes, we wanted to explore users' opinions on adding citations to YouTube. There was also a distinction made between informational and entertainment videos. Participants were unsure about the effectiveness that citations would have on entertainment videos, and some, like P8, noted that citations would be more useful on informational videos. ``\textit{A lot of the time, I’m watching [YouTube videos] for entertainment purposes, but for a more informational video, I would definitely want to see [citations] throughout the video and on the screen or on the side}''. Multiple participants mentioned that they would want to see citations on topics such as medicine, history, and politics, or ``\textit{topics with important life consequences}'' (P10).
Another instance that participants mentioned they would like to see citations is on videos that present an opinion they disagree with. As P8 said, ``\textit{I think a video like this where I don’t really agree with the opinion of, I would want to explore [the citations] to like, see what their perspective is and know what their thought process was}''. In this case, the citations could help the user better understand an opposing perspective. A few participants also mentioned that they would want to be able to ‘flag’ parts of videos, especially when only certain parts of a video contain false information.
There was some disagreement among participants about whether increasing the number of citations on a video would make it more credible or less credible. One participant thought that adding citations could show that the content creator did their research. However, as P7 said, ``\textit{I feel like, the more I talk about this, the more I feel like [seeing citations] might be helpful because if you see a lot of citations on a video it might be a red flag to be like, oh okay maybe I should take a closer look at this}''. They thought that seeing a video with many citations could be a sign that there is a lot of disputed information in the video. In adding citations, one way this issue could be addressed is through providing multiple citation types such that disputed information could be distinguished from supporting information.

\section{Viblio}\label{viblio}
From our exploratory interviews, we found that people make use of multiple existing signals on video-sharing platforms to determine credibility. However, their heuristics for credibility included outright incorrect ones, such as incorrectly assuming information panels serve as endorsements, as well as ones that may lead them astray, such as the verification signal or production quality. We also took inspiration for how to implement credibility signals from other major information sources online, like citations (Wikipedia), knowledge panels (Google), and crowdsourced misinformation warnings (Community Notes).
With these findings in mind, we developed a prototype system called Viblio. Instantiated as a Chrome extension for YouTube, Viblio allows users to add and view crowdsourced citations on any YouTube video.

\subsection{Design}\label{viblioDesign}

Prior to implementing our system, we developed some design goals for Viblio based on the findings of our exploratory study, other credibility-focused systems, and extant literature. For functionality, the system should allow anyone to add a citation and view all citations on any YouTube video. We wanted to add granularity to these citations by only displaying them for a certain time frame, as perhaps only a part of the video requires a citation. Feedback from interviewees regarding their usage of citations showed that it was important to specify the purpose of a citation, such as whether it supports or refutes content in the video or simply provides further information. Participants also expressed that they would only want or need citations in specific circumstances, so we wanted the system to be low-profile when not in use.

As we designed Viblio, we kept our findings from the exploratory study and prior work in mind. Community Notes's initiative that allows users to identify and annotate tweets that they believe contain misinformation or false content inspired our use of a crowdsourcing approach for the collection of citations \cite{Introduc53:online}. Based on extant literature (\cite{sundar2008cognitive}), we wanted to activate users' agency in shaping their perception of content credibility. It was important for us to be able to display information from experts, reputable organizations, or authoritative accounts, as users are inclined to trust this content more \cite{briggs2002trust,lin2016social}.

Our exploratory study shows that existing credibility signals can play a large role in users' credibility judgment. We thought it was particularly interesting how users went to look at the video description on the playback page, which influenced our decision to place the extension below the video. It was also surprising how easily signals could get misinterpreted, like YouTube's information panels on misinformation-prone topics. As a result, when designing our system, we wanted to make it clear that the extension was a separate add-on to the video that referenced citations and not any sort of endorsement by outside sources. Although we purposefully kept the citations separate, we also did not want the interface to distract or interfere with the user’s experience watching videos. As P10 mentioned in \ref{citationquote}, we wanted our system to be able to blend into the interface when not needed. For this reason, the elements, colors, and shapes within the extension closely mirror YouTube’s existing interface.

Another important factor we considered in the design process was indicating the citation type. It was important for users to be able to quickly identify whether a citation supported or refuted the information in the video. Based on suggestions from participants, we also chose to include a third category of citations, extra information. This type is intended for sources that provide further context or expand upon the information provided within the video. We also wanted to be careful about how this labeling would affect the viewer's overall opinion, especially as recent work found that stance-based labels may intensify selective exposure and lead to users being more vulnerable to polarised opinions and fake news \cite{10.1145/3274324}.

When considering the permission-levels and potential for misuse of Viblio, we wanted to focus first on whether citations made in good faith would have an impact or have any utility for users and how best to present citations. After validating this, future work could consider how to guard against bad-faith actions. We discuss possibilities in Section \ref{future-work} (Future Work) that draw ideas from successfully moderated crowd-powered systems.

\subsection{System Description}

\begin{figure*}
\centering
\includegraphics[width=1\textwidth]{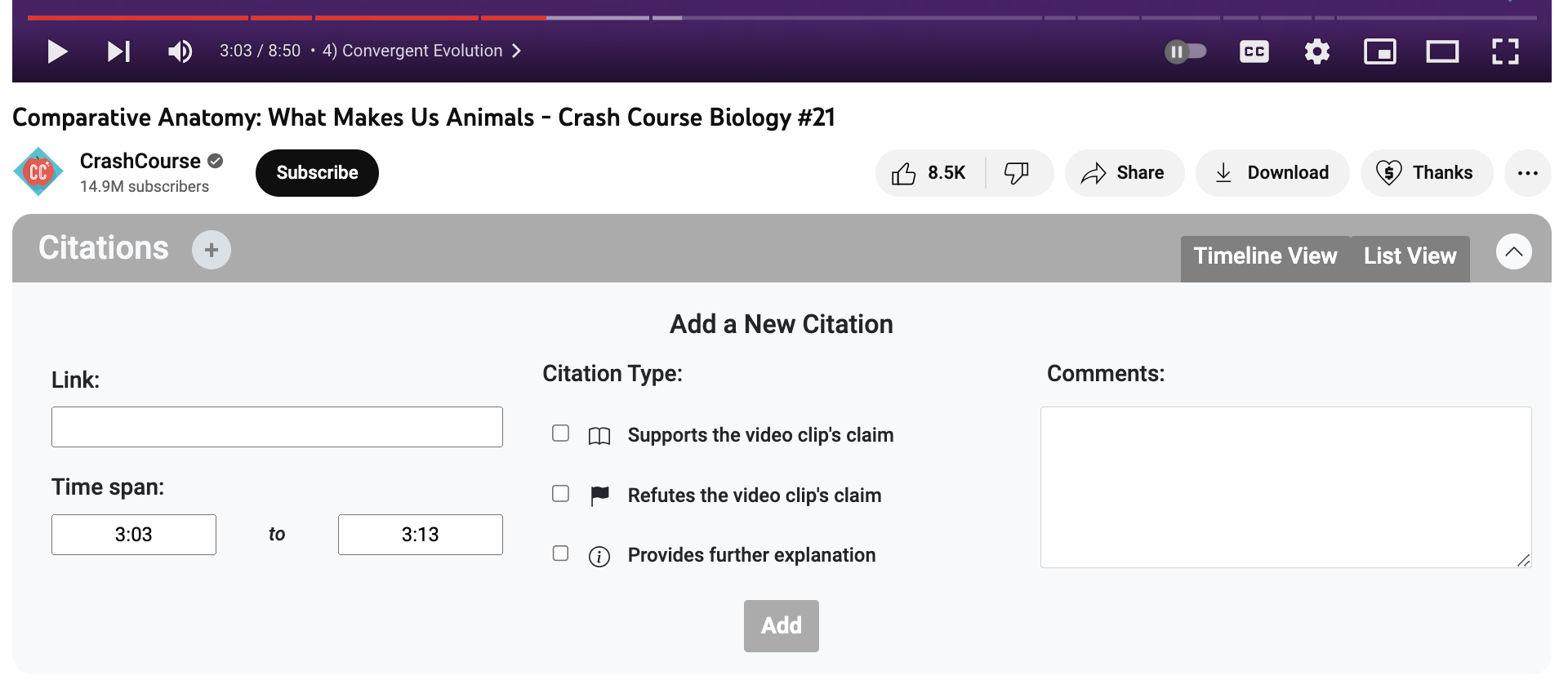}
\subfloat[\label{add-view} Add View]{\hspace{1\linewidth}}

\hfill\newline

\includegraphics[width=1\textwidth]{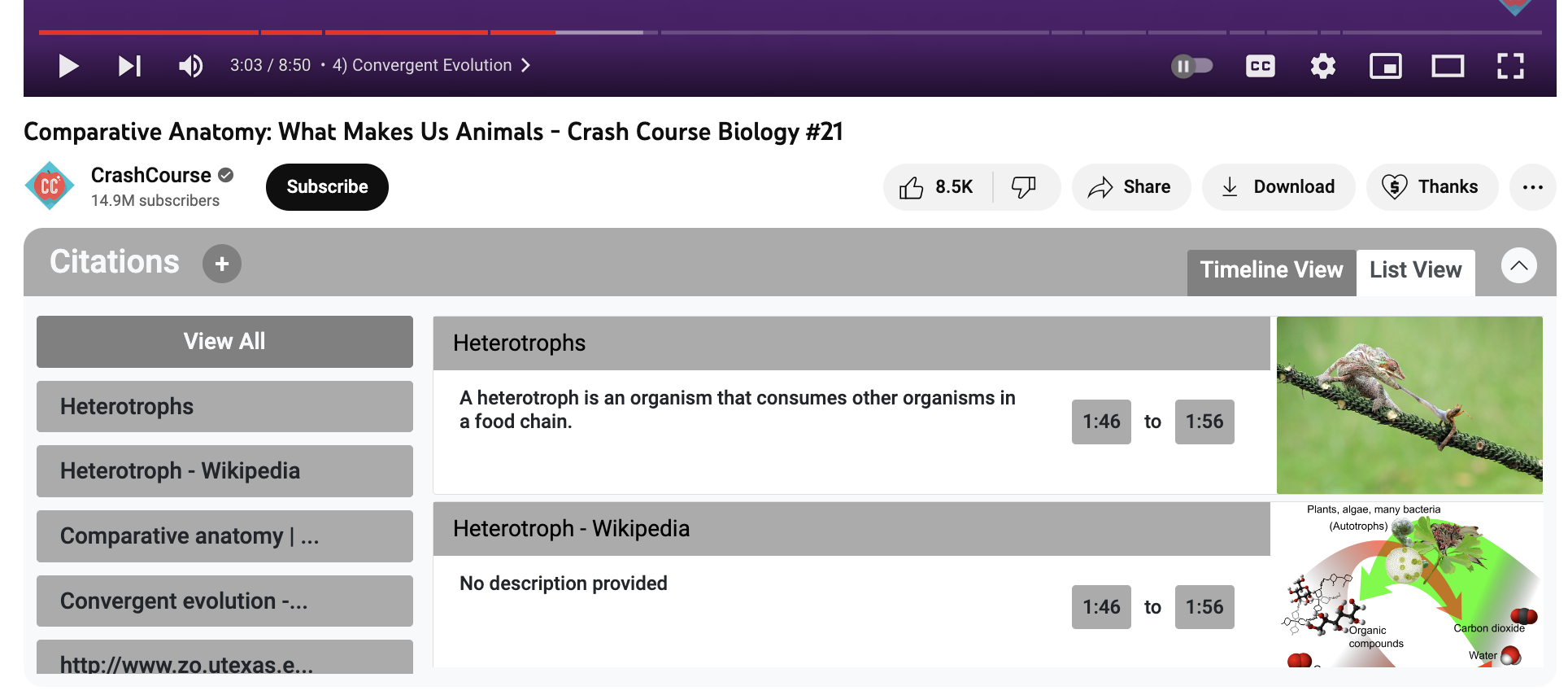}
\subfloat[\label{list-view} List View]{\hspace{1\linewidth}}
\caption{Viblio's list view and add-citation view. Add-citation view shows where participants enter relevant data for a new citation. In list view, shortcut buttons for each citation are on the left, while citations are shown on the right.}
\end{figure*}

\paragraph{Citation insertion} When the user wants to add a citation, they can click the ``Add citation'' button on the extension. This opens up a form where users can input citation information such as the link to the citation, the type of citation, any comments or notes, and start/end times for the section of the video it references (Fig. \ref{add-view}). The start time is automatically populated to the current video playback time, and the end time is 10 seconds later than the start time (or the end of the video, if that is sooner). Both fields can be adjusted by the user. On the backend, information about the source is scraped from the provided link, such as the title, author, source, description, and cover image. 

\paragraph{Timeline view} The default view on Viblio is the timeline view (Fig. \ref{timeline-view}). A timeline is displayed across the top of the Viblio window, which syncs with the video timeline. If a user jumps ahead in the video, the Viblio timeline will mirror this. Any crowdsourced citations added to the video will appear as circles along the timeline. As the YouTube video plays, once the video enters a time frame specified by a citation creator, the relevant citation will be displayed under Viblio's timeline. Users can also navigate to the start time of the video clip associated with a citation by clicking on the corresponding circle on Viblio's timeline. 

\paragraph{List view} The other way to explore citations is through the list view (Fig. \ref{list-view}). By default, all of the video citations are displayed in a scrollable list. Shortcut buttons to individual citations are shown on the left side of the window. The citations on this page are not displayed based on when their specified time frame is currently being played, so they give the user a chance to explore all of the citations in depth.

\paragraph{Citation types} From our contextual inquiry, we learned that users may use citations for a variety of purposes, such as references or a bibliography, flags for misinformation, and to provide extra information. As a result, we wanted to incorporate a way to represent these different types of citations in the system. We included options to ``support the video clip claim'', ``refute the video clip claim'', and ``provide further explanation'' (as seen in Fig. \ref{add-view}). Depending on the type of citation, an icon is displayed on the citation, and a color treatment (green for supportive, red for refuting, or blue for informative) is used throughout the extension in reference to the specific citation, such as in the citation's coordinating circle on the timeline. Using colors and icons to indicate the citation type allows users to quickly identify the relationship of the citation to the video.

\paragraph{Citation format} When a citation is displayed within Viblio, a consistent format is used. We based the information collected and displayed for each citation on popular citation formats and the norms set forth by Wikipedia. Depending on the information that was able to be scraped from the provided link, the citation block will display the title, source, description, and cover photo. Along with the metadata, any comments or notes the user made when adding the citation, the timespan the citation is referenced in the video, and the type of citation is displayed.

\subsection{System Implementation and Considerations}

This extension was implemented for Chrome using JavaScript (React). The back end is an Azure CosmosDB database that interfaces with a React front end through a Python Flask API deployed as an Azure Web App. We chose to develop a Chrome extension for YouTube over a separate video-playback platform as our goal was to build upon users’ existing behaviors and an established social platform. 

However, this decision also required some trade-offs due to the limitations of developing a Chrome extension. The main trade-off we made was that Viblio would only be able to be used on a computer and would not be able to be used on the YouTube app or mobile Chrome. A more integrated approach would allow Viblio to be used on the app or mobile devices, responsible for around 90\% of YouTube’s watch time  \cite{noauthor_youtube_2023}. In the meantime, our prototype allows us to quickly experiment with and deploy new credibility signals in real-world settings.

\section{Reflective Study Overview}\label{reflective-study}

In order to evaluate our system's efficacy as an additional credibility signal on YouTube, we deployed Viblio in an extended user study. Participants were tasked with using Viblio on a given set of videos over the course of two weeks. Our goal with this study was to evaluate Viblio as a new credibility signal, Viblio's use in the context of other credibility signals, Viblio's effect on participants, and the topics for which Viblio could be of use to YouTube users.

\subsection{Methods}
Participants were primarily recruited through word-of-mouth and X and were asked to provide demographic information, including age range, gender, political preference, and education level (Table \ref{reflective-participants}). They were also asked about their familiarity with YouTube, the type of content they typically watched, and the average amount of time they spent on YouTube (hours/week). Responses to recruiting materials were then screened for comfort with screen sharing, being recorded in online meetings, comfort using Chrome, and basic knowledge using YouTube. Generally, participants were selected by sign-up order with an aim to also diversify the participant group as much as possible. Of the 12 participants, half were in the 23-30 age range, and three were over 45. Overall, the participants skewed male (7/12), educated (7 with Master’s, 5 with Bachelor’s), and usually spent up to 7 hours on YouTube a week (6 spent less than 3, 5 spent 3 to 7 hours). The group was relatively politically diverse, with 3 identifying as Democrats or left-leaning, 2 identifying as Moderate or centrist, 1 identifying as Republican or right-leaning, 1 identifying as independent, and 4 choosing not to share. Upon completion of the entire study, participants were compensated with a \$60 gift card. One participant (P6) did not fully complete the study; however, his responses were included for the exercises he did complete.

\begin{table*}[t]
\small
\centering
\caption{Participant Demographics in the Reflective Study}
\begin{tabular}{c|c|c|c|c|c|c}
\hline
\label{reflective-participants}Participant & Age   & Gender & Education Level          & Political Preference                                                             & \begin{tabular}[c]{@{}c@{}}Average Hours \\ On YouTube \\ Per Week\end{tabular} & \begin{tabular}[c]{@{}c@{}}Primary Use \\ for YouTube\end{tabular} \\ \hline
1           & 23-30 & Male   & Master's Degree or above & Chose not to share                                                               & 3-7                              & Entertainment           \\
2           & 18-22 & Male   & Bachelor's Degree        & Democrat/Left-leaning                                                            & 3-7                              & Entertainment           \\
3           & 45+   & Female & Bachelor's Degree        & Democrat/Left-leaning                                                            & \textless{}3                     & Entertainment           \\
4           & 45+   & Male   & Bachelor's Degree        & Moderate/centrist                                                                & 3-7                              & Education               \\
5           & 45+   & Male   & Master's Degree or above & Republican/right-leaning                                                         & \textless{}3                     & Current events/news     \\
6           & 31-45 & Male   & Master's Degree or above & Moderate/centrist                                                                & \textless{}3                     & Education               \\
7           & 23-30 & Male   & Master's Degree or above & Independent & 3-7                              & Entertainment           \\
8           & 23-30 & Female & Master's Degree or above & Unsure                                                                           & 3-7                              & Entertainment           \\
9           & 23-30 & Female & Master's Degree or above & Democrat/Left-leaning                                                            & 7+                               & Entertainment           \\
10          & 18-22 & Female & Bachelor's Degree        & Chose not to share                                                               & 3-7                              & Entertainment           \\
11          & 23-30 & Male   & Bachelor's Degree        & Chose not to share                                                               & \textless{}3                     & Entertainment           \\
12          & 23-30 & Female & Master's Degree or above & Chose not to share                                                               & \textless{}3                     & Entertainment           \\ \hline
\end{tabular}
\end{table*}

The study took place virtually on Zoom over a period of around 4 weeks, although each participant took a maximum of two weeks individually to complete the study. Each participant made their way through three distinct sections of the study: an onboarding interview, independent activities using Viblio, and a reflective interview.

\paragraph{On-boarding interview} For the on-boarding interview, we started by assisting the participants through downloading and installing Viblio. Depending on the participant’s technology experience, this sometimes involved screen-sharing to provide a visual aid. Once Viblio was installed, participants were given the link to an onboarding--specific YouTube video. Participants were then given time to explore and interact with Viblio. Once they felt comfortable with the extension, we then led participants through usability exercises. These exercises involved navigating to the Timeline and List views, viewing specific citations, and adding a new citation to the video. For the purpose of the usability test, the content of the citation added was not important. Once the onboarding and usability exercises were complete, any remaining time was spent ensuring that participants fully understood how to use Viblio before the independent exercises. 

\paragraph{Independent exercises} The main portion of this study is the independent exercises each participant completed. Over the course of around 15 days, each participant was given a schedule of videos to watch, split across ten days. Each day, participants had 2--3 videos to watch (totaling under 30 minutes) and a reflective survey to complete. Participants were divided into three randomized groups, and each group received a unique schedule. The schedules were planned so that each group would encounter videos at different stages. Some videos would not have any citations added, while others would already be populated with citations by other participants. This method ensured that participants encountered a variety of citation volumes across videos.

\paragraph{Daily survey} For each day of the study, each participant would fill out a daily survey for the videos they watched. Participants were asked to fill out this survey while completing the exercises so that responses would be of the moment. Before watching each video and interacting with Viblio, the participant would rate how credible they believed the video was on a scale of 1--5 and explain their reasoning behind the rating. Once the participants watched the video and interacted with Viblio, they would then rate the credibility again. If their score changed, they were asked to explain what had affected their rating (or not). If the participant came across something in the video that they believed needed a citation and there was not one already added by another participant, we asked them to find an appropriate source and add a citation. Participants would also record this event in the daily survey. 

\paragraph{Video set} We developed the set of videos used in the study to cover a variety of topics, types of video, possible source types, use cases, and credibility levels (Table \ref{reflective-videos}). The topics covered by the videos went from informative topics like the history of cereal and laser eye surgery, to controversial subject matters like election fraud, COVID-19 vaccines, and abortion. The types of video included educational, entertainment-based, and news reporting. Videos were published from well-known sources like Fox News and Crash Course, as well as from small local news stations and independent creators. A few of the videos purposefully contained problematic or false information. During the reflective interview, participants were informed of the videos, including misinformation.

\paragraph{Reflective interview} Once each participant finished the independent exercises, a semi--structured final reflective interview was scheduled. The first part of the interview focused on the participant’s experience throughout the study. Questions were asked about the usability, their process for exploring and adding citations, how they chose whether or not to add a citation, how Viblio affected their YouTube experience, how Viblio affected their credibility ratings, and the quality of citations they came across. The participant’s responses to the daily survey were reviewed beforehand, then any outliers or noteworthy responses were discussed. This discussion helped clarify motivations, the participant’s process, and any confusion. The last topic of the interview covered possible applications of the system. Participants were asked to evaluate if they could see themselves using this system in the future and, if so, where they would use it.

\paragraph{Analysis}The three sections of the study were analyzed individually. Open-coding methods were used to complete a sample set from each section. Once a code-book was established, the three data sets, the transcripts from the onboarding and reflective interviews and the daily survey, were coded. Based on the code book, emerging topics were identified by thematic analysis. Some of the codes included but were not limited to design feedback, usability, use case, citation quality, and process. Throughout the study, data was also recorded whenever users interacted with Viblio and the citations. Given the small sample size, the statistical results of this data collection are not significant to our findings. We do, however, refer to this data to discuss participant's usage of Viblio throughout the study. Usage data results can be viewed in Tables \ref{appTable1} and \ref{appTable2} in the Appendix.

\subsection{Results}
\subsubsection{Impact of Citations as a Credibility Signal}
Analysis of the daily participant survey highlights the dynamics of credibility assessment in the context of various video genres, shedding light on the intricate interplay between source credibility, the presence of citations, and the influence of individual perceptions and prior beliefs. While some participants found citations instrumental in confirming their skepticism regarding untrustworthy statements from known sources, others found citations to be essential in mitigating biases, especially in videos that lacked alternative viewpoints. 
Our exploratory study showed that signals such as the video content, source reputation, and sensationalism influenced participants' credibility assessments. Citations served as an additional credibility signal in these cases, reinforcing trust in familiar and reputable sources. 

\paragraph{Evaluating news sources and political content}
In the evaluations of videos with political content, participants placed significant emphasis on source credibility. Established sources like NBC News and Reuters were generally perceived as more trustworthy. However, pre-existing perceptions of bias in news sources, such as Fox News, influenced participants' initial credibility ratings. The presence of citations within these videos played a crucial role in enhancing post-watch credibility, particularly by validating the claims made in the video. In a video published by Fox News, P11 first commented, ``\textit{Fox News is mostly credible, but sometimes also spreads misinformation}''. After exploring the citations, P11 raised his credibility score and commented, ``\textit{I think the citations helped me validate the facts in the video and hence I increased my credibility score}''.

Participants found comfort in citations from reputable sources and the inclusion of direct video footage or quotes, such as in a video focused on a strike at Rutgers University. Participants appreciated the lack of commentary or controversial assertions in the video, which contributed to its credibility. While some participants did not add citations because the video did not make specific claims, those who did found that the citations provided additional context and information about the strike, reinforcing the video's credibility. 

Videos discussing legal cases and investigations received mixed credibility ratings. Participants considered the nature of the topic and the source's reputation when evaluating credibility. Citations were valuable in providing additional context, fact-checking, and validation of claims, which contributed to increased credibility. Skepticism was often related to concerns about the evidence presented in the videos and the potential for bias. For example, in a video investigating ExxonMobil, one participant found that ``\textit{at its first appearance the video seemed to be a direct attack on ‘ExxonMobil’, though there are multiple other industries and factors that are responsible for the increased plastic production and pollution. Not a clear mention of this made me question the credibility of the video as it tried to overshadow the actual issue at hand}'' (P12). After reviewing the citations, P12 found that ``\textit{some of the citations actually backed the claims or provided neutral evidence, removing the bias that I earlier had}''. 

\paragraph{Assessing scientific and educational content}
Videos featuring educational and scientific content from trusted sources like CrashCourse and TED-Ed were consistently perceived as highly credible. Participants' familiarity with the source and previous positive experiences with informative, unbiased content influenced their pre-watch credibility ratings. On a CrashCourse video, P11 noted: ``\textit{I was familiar with the content discussed in the video and could verify its credibility. Also, I have personally watched a lot of videos created by this channel and can vouch for its content (based on my prior experience)}''. While citations were present in these videos, they were viewed as less crucial, as participants deemed the content itself well-researched and trustworthy. In this case, the existing credibility signals proved to be enough for participants and citations were not used as a credibility signal. Instead, citations primarily played a role in providing additional context and facilitating deeper exploration of the topics discussed. On a TED-Ed video, P6 commented: ``\textit{I liked the citations because they allowed me to quickly explore topics that were mentioned in the video}''; whilst on a CrashCourse video, P12 commented: ``\textit{the citations helped me read up about the things discussed in the video in detail, increasing my notion of credibility for the video}''. Sources in the citations of these videos were generally clicked on more than average (see Table \ref{appTable1}). Trust in the source, informative content, clear explanations, and visual evidence through experiments contributed to participants' positive credibility assessments.

\paragraph{Evaluating videos on sensitive and controversial topics}
Videos discussing sensitive and controversial topics, such as abortion and election fraud, often faced credibility challenges. Participants expressed skepticism based on concerns about potential misinformation and bias in these videos. In a Joe Rogan video surrounding COVID-19 vaccine controversy (Video 22), P9 noted: ``\textit{I have preconceived notions that discredit Joe Rogan. Also, I found that the sensationalization about the title and the intro reduced my feelings of credibility about the video}''. In such cases, citations played a crucial role in influencing credibility by corroborating or refuting the video's claims. Participants relied on citations to fact-check claims and provide additional context, which significantly impacted their overall credibility assessments. On the same video, P11 commented: ``\textit{The citations helped me validate some of the facts in the video, and also helped refute some of the misinformation in the video}''. Skepticism was often linked to the controversial nature of the topic, and participants highlighted the need for further verification and more objective reporting.

\hfill\newline
\indent In summary, participants' credibility assessments were initially shaped by factors and existing signals such as the perceived credibility of the news source, the sensitivity of the topic, and the clarity of presentation. The presence and relevance of citations played a critical role as a credibility signal in either strengthening or weakening the videos' credibility, depending on their alignment with the video's content and participants' preconceptions. The influence of participants' prior knowledge and biases about specific topics and sources also significantly shaped their credibility evaluations.

\begin{table*}[!htbp]
\vspace{-5pt}
\footnotesize
\centering
\caption{Videos used in Reflective Study with participant ratings.}
\resizebox{!}{1.25\textwidth}{%
\begin{sideways}
\begin{tabular}{c|c|c|c|c|c|c|c}
\hline
\label{reflective-videos}\rotatebox{0}{\textbf{Video}} & \rotatebox{0}{\textbf{Title}}                                                                                                                                 & \rotatebox{0}{\textbf{Source}}                                                       & \rotatebox{0}{\textbf{Area Tags}}                                                                   & \textbf{\rotatebox{0}{\begin{tabular}[l]{@{}l@{}}Participant\\ Responses\end{tabular}}} & \textbf{\rotatebox{0}{\begin{tabular}[l]{@{}l@{}} Initial Credibility  \\ Score Average \end{tabular}}} & \textbf{\rotatebox{0}{\begin{tabular}[l]{@{}l@{}}Final Credibility\\ Score Average\end{tabular}}} & \rotatebox{0}{\textbf{Difference}} \\ \hline
1              & \begin{tabular}[c]{@{}c@{}}``COUNT ONLY LEGAL VOTES'' Rudy Giuliani \\ OPENING STATEMENT On Election Fraud Claims\end{tabular}                   & Fox News                                                              & \begin{tabular}[c]{@{}c@{}}Political, News, \\ Controversial\end{tabular}            & 9                                                                        & 2.78                                                                                 & 2.56                                                                               & -0.22               \\
2              & Biden says he's 'planning on running in 2024'                                                                                                  & Fox News                                                              & Political, News                                                                      & 10                                                                       & 3.10                                                                                 & 3.70                                                                               & 0.60                \\
3              & Biden: ‘I plan on running’ in 2024 presidential election                                                                                       & NBC News                                                              & Political, News                                                                      & 12                                                                       & 4.00                                                                                 & 4.50                                                                               & 0.50                \\
4              & \begin{tabular}[c]{@{}c@{}}California investigates ExxonMobil over its \\ plastics waste\end{tabular}                                          & \begin{tabular}[c]{@{}c@{}}KPIX | CBS NEWS \\ BAY AREA\end{tabular}   & Political, News                                                                      & 12                                                                       & 3.08                                                                                 & 4.00                                                                               & 0.92                \\
5              & \begin{tabular}[c]{@{}c@{}}Comparative Anatomy: What Makes Us Animals - \\ Crash Course Biology \#21\end{tabular}                              & CrashCourse                                                           & Educational, Scientific                                                              & 11                                                                       & 4.27                                                                                 & 4.82                                                                               & 0.55                \\
6              & Deadline on abortion pill ruling                                                                                                               & 5NEWS                                                                 & \begin{tabular}[c]{@{}c@{}}News, Controversial, \\ Sensitive\end{tabular}            & 10                                                                       & 3.40                                                                                 & 3.80                                                                               & 0.40                \\
7              & Fake heiress Anna Sorokin to be deported l GMA                                                                                                 & \begin{tabular}[c]{@{}c@{}}Good Morning \\ America (GMA)\end{tabular} & News                                                                                 & 12                                                                       & 3.58                                                                                 & 4.08                                                                               & 0.50                \\
8              & Fox settles Dominion case, but bigger lawsuit looms                                                                                            & Reuters                                                               & Political, News                                                                      & 11                                                                       & 3.82                                                                                 & 4.27                                                                               & 0.45                \\
9              & GOP ban on trans student athletes passes House                                                                                                 & Reuters                                                               & \begin{tabular}[c]{@{}c@{}}News, Controversial, \\ Sensitive\end{tabular}            & 10                                                                       & 4.10                                                                                 & 4.40                                                                               & 0.30                \\
10             & Heredity: Crash Course Biology \#9                                                                                                             & CrashCourse                                                           & Educational, Scientific                                                              & 11                                                                       & 4.00                                                                                 & 4.09                                                                               & 0.09                \\
11             & How does laser eye surgery work? - Dan Reinstein                                                                                               & TED-Ed                                                                & Educational, Scientific                                                              & 12                                                                       & 4.58                                                                                 & 4.58                                                                               & 0.00                \\
12             & \begin{tabular}[c]{@{}c@{}}Ilhan Omar connected Ballot Harvester in cash-for-ballots \\ scheme: ``Car is full'' of absentee ballots\end{tabular} & Project Veritas                                                       & \begin{tabular}[c]{@{}c@{}}Political, News, \\ Controversial\end{tabular}            & 8                                                                        & 2.50                                                                                 & 1.88                                                                               & -0.63               \\
13             & \begin{tabular}[c]{@{}c@{}}Joe Biden’s America shouldn’t be our future: \\ Matthew Whitaker\end{tabular}                                       & Fox News                                                              & Political, Controversial                                                             & 11                                                                       & 2.36                                                                                 & 2.09                                                                               & -0.27               \\
14             & \begin{tabular}[c]{@{}c@{}}Kentucky Gov. Beshear lost 'close friend' in \\ Louisville shooting\end{tabular}                                    & Fox News                                                              & Political, News                                                                      & 12                                                                       & 3.42                                                                                 & 4.00                                                                               & 0.58                \\
15             & \begin{tabular}[c]{@{}c@{}}Leak of classified documents and intelligence shakes \\ U.S. Department of Defense\end{tabular}                     & Yahoo Finance                                                         & \begin{tabular}[c]{@{}c@{}}Political, News, \\ Controversial\end{tabular}            & 11                                                                       & 3.09                                                                                 & 3.73                                                                               & 0.64                \\
16             & \begin{tabular}[c]{@{}c@{}}Major anti-abortion group gives scathing response \\ to Trump comments\end{tabular}                                 & MSNBC                                                                 & \begin{tabular}[c]{@{}c@{}}Political, News, \\ Controversial, Sensitive\end{tabular} & 8                                                                        & 3.38                                                                                 & 4.25                                                                               & 0.88                \\
17             & mRNA vaccines, explained                                                                                                                       & Vox                                                                   & Educational, Scientific                                                              & 12                                                                       & 3.92                                                                                 & 4.50                                                                               & 0.58                \\
18             & \begin{tabular}[c]{@{}c@{}}Picket lines form at Rutgers University campuses \\ during historic strike\end{tabular}                             & NJ.com                                                          & News                                                                                 & 10                                                                       & 4.00                                                                                 & 4.70                                                                               & 0.70                \\
19             & \begin{tabular}[c]{@{}c@{}}Ruling Against Abortion Pill Mifepristone Could \\ Fundamentally Alter The FDA, Becerra Warns\end{tabular}          & \begin{tabular}[c]{@{}c@{}}Forbes Breaking \\ News\end{tabular}       & \begin{tabular}[c]{@{}c@{}}Political, News, \\ Controversial, Sensitive\end{tabular} & 12                                                                       & 3.50                                                                                 & 4.50                                                                               & 1.00                \\
20             & \begin{tabular}[c]{@{}c@{}}US secrets leaked in social media post containing \\ Ukraine, Russia documents\end{tabular}                         & Fox News                                                              & Political, News                                                                      & 12                                                                       & 2.67                                                                                 & 3.08                                                                               & 0.42                \\
21             & \begin{tabular}[c]{@{}c@{}}WH 'disagrees strenuously' with Texas judge's decision \\ on abortion pill\end{tabular}                             & MSNBC                                                                 & \begin{tabular}[c]{@{}c@{}}Political, News, \\ Controversial, Sensitive\end{tabular} & 10                                                                       & 3.40                                                                                 & 4.20                                                                               & 0.80                \\
22             & \begin{tabular}[c]{@{}c@{}}What Did Bill Gates Say About COVID Vaccine \\ Side Effects?\end{tabular}                                           & \begin{tabular}[c]{@{}c@{}}PowerfulJRE \\ (Joe Rogan)\end{tabular}    & \begin{tabular}[c]{@{}c@{}}Political, News, \\ Controversial\end{tabular}            & 8                                                                        & 2.50                                                                                 & 2.50                                                                               & 0.00                \\
23             & When a physics teacher knows his stuff !!                                                                                                      & \begin{tabular}[c]{@{}c@{}}Lectures by \\ Walter Lewin\end{tabular}   & Educational, Scientific                                                              & 11                                                                       & 4.73                                                                                 & 5.00                                                                               & 0.27                \\
24             & Is Organic Food Worse For You?                                                                                                                 & ASAPScience                                                           & Educational                                                                          & 5                                                                        & 3.60                                                                                 & 4.20                                                                               & 0.60                \\
25             & Cereal - A Brief History - The New Yorker                                                                                                      & The New Yorker                                                        & Educational                                                                          & 5                                                                        & 3.40                                                                                 & 3.80                                                                               & 0.40                \\ \hline
\end{tabular}
\end{sideways}}
\vspace{-25pt}
\end{table*}

\subsubsection{Viblio's Use as a Credibility Signal}\label{useascred}
In this study, participants offered insight into their experiences with Viblio, a tool designed to intentionally center credibility on YouTube. Their feedback ranged from usability strengths and concerns to motivations for engaging with citations and perceptions of their impact alongside existing signals. Participants demonstrated diverse strategies for interacting with citations and varied opinions on their utility. Some envisaged Viblio's potential in different contexts, while others highlighted factors influencing changes in credibility scores. Their willingness to use Viblio in the future was shaped by context and personal preferences. 

\paragraph{Usability and design} Participants provided valuable feedback on the usability of Viblio, highlighting the overall utility and areas for potential improvements. Every participant found the extension to be intuitive and easy to use, with P5 even stating: ``\textit{I would be an enthusiastic yes to advocate for YouTube to add this functionality natively}''. The main feedback we received related to Viblio’s integration into the YouTube interface. Some users (P1, P4, P7, P10, P12) found the division between the video itself and the extension hard to bridge. P4 and P11 expressed interest in an interface that overlays the video for a more integrated approach. P10 found scrolling between the video and the extension distracting. P7 discussed how it was difficult to process the video content and read the citations at the same time: ``\textit{I think, usually, like, my YouTube routine is a lot more, like, I guess less cognitively intensive… I don't think twice or not whether or not this is true}''.

\paragraph{Engagement process} Participants engaged with Viblio's citations using diverse strategies, reflecting their individual information-seeking habits and Viblio’s flexibility to accommodate various user preferences. The majority of participants would use the timeline view while simultaneously watching the video. A few participants (P2, P10, P11) shared that they would actively pause the video and explore the citations when unfamiliar topics came up. P1 and P3 used Viblio to ‘fact check’ when questioning information. One participant (P7) preferred to take the time to explore citations before watching a video. In general, participants would use the timeline view while the video was playing but would use the list view when actively engaging with and exploring the citations. For example, P9 shared that she would use List View more: ``\textit{I use the list format more than the timeline format and honestly, usually I'd, like, click the video and then kind of scroll through the list and it's almost like a little outline of the video… I thought it was really helpful}''. In Table \ref{appTable2}, our interaction data shows how the majority of participants had a primary view that they used to interact with citations. Another interesting takeaway from our usage data involves the frequency at which participants practiced lateral reading. The majority, nine, of our participants clicked through to an article between 0 and 5 times. Three of our users, however, explored a citation's source more: 10, 20, and 35 times. 

\paragraph{Motivations for adding citations} Participants in the Viblio platform exhibited a wide range of motivations for adding citations during the study. An overarching theme displayed by many participants was the motive to add citations when encountering ambiguous or untrue information. One participant cited that her motivation lay in curiosity and the urge to fact-check questionable or disagreeable content: ``\textit{I think it was that curiosity, or that, like, 'is that right?' that was when it was easiest to make the citation}'' (P9). Four other participants (P1, P3, P10, and P11) expressed similar sentiments. Three participants, P4, P5, and P7, approached adding citations through a balance-based perspective, especially in political discussions. P5, for instance, emphasized the importance of reinforcing or counteracting points within the video, stating, ``\textit{If there's something that I'm aware of from my knowledge that reinforces a particular point or counters a particular point, then I would take the time to research and post it}''.  The third main motivation we identified in the participant's responses was to seek out information gaps where citations could be valuable. Participants who expressed this as their primary motivation (P1, P2, and P12) also highly valued citations of good quality and credibility benefits.

\paragraph{Impact and utility of citations} Participants' perceptions of the impact and utility of citations varied across topics and applications, but all participants noted at least one case in which they were helpful. On the other hand, participants found citations to be of limited impact or redundant in cases where they had prior knowledge. In contrast, multiple participants emphasized their value in relation to unfamiliar video content. For example, P5 commented: ``\textit{I would be more likely to sway my opinion on things that I either didn't have familiarity with, or didn't feel strongly about one way or the other. Then I would be more open to changing views or deeper understanding}''. P2 found that citations helped to mitigate some of his initial bias on a source, improving his final credibility score. Citations from surface-level or introductory sources, such as Wikipedia, were often seen as redundant.

Another area that participants found the citations particularly helpful in determining credibility was for video content they personally disagreed with or were uncertain about. P9 stated: "\textit{Honestly, I think I found [citations] most useful when they disagree with the video content… And second most useful when it was about a topic I wasn't sure about}''.

In educational or informative videos, citations primarily played a different role than credibility. Many participants shared that they found most educational content trustworthy, as it is more normal for these content creators to verify their credibility by sharing where their information comes from. In this case, Viblio became a learning tool and a jumping-off point for further exploration. Citations throughout the video provided an easy way for participants to engage with related materials.

\paragraph{Utility of Viblio in various contexts} Participants expressed various views on using Viblio in different settings, especially when related to controversial or political topics. All participants agreed upon the benefits and value of using Viblio in scientific videos and fact-based contexts. P10 highlighted Viblio's potential in medical information videos, where accurate facts are crucial. Participants also found the citations not to be useful in entertainment-based content. This was generally attributed to a lack of desire to determine a video’s credibility when watching for entertainment primarily.

Participants’ views on Viblio’s use in political or controversial topics presented the widest variety of standpoints. While some participants believed that Viblio would be essential to controversial topics, others expressed reservations about misuse. In Viblio’s current form, P5 expressed concern that Viblio could potentially lead to ``\textit{wars of refutations}'' if not used judiciously. Other participants also expressed concerns about political video citations potentially fueling disputes between political groups. On a more personal level, participants were also unsure how they would interpret citations on political videos: ``\textit{This part of me thinks that the more political content, I would use it more for, and part of me also feels like I wouldn't trust a lot of the citations that came up on that, because you can find a headline supporting almost anything in my perspective}'' (P9). Looking beyond the concern, some participants envisioned Viblio as a tool to balance perspectives within news videos and facilitate neutral conversations.

The feedback and experiences shared by participants in this study shed light on the complexities of incorporating citation features into online video platforms like YouTube. Users expressed a range of preferences, motivations, and perceptions regarding Viblio's usability and impact on their viewing experience. While some participants embraced the tool's potential in specific contexts, others called for improvements to enhance cohesion and detail in Viblio.

\section{Discussion}\label{discussion}
The introduction of Viblio as an extension to YouTube has shown a promising shift in how users evaluate the credibility of online video content. Our evaluative study, involving participants with diverse information-seeking habits and preferences, revealed a remarkable level of agency granted to users through Viblio. While participants held a wide array of opinions regarding the contexts in which Viblio would prove effective, a unanimous consensus emerged: Viblio offered a means for users to take control of their credibility assessments. In this discussion section, we delve into the implications of these findings and elucidate the broader applications and potential ramifications of Viblio.

\subsection{Empowering User Agency and the Role of Citations}

One of the key takeaways from our study is the extent to which Viblio empowers users to form their own opinions about video content. Participants emphasized that Viblio's value extended beyond the mere presence of citations; rather, it provided them with a crucial evaluative point alongside existing signals in assessing credibility. Despite the value users mentioned, it was apparent that Viblio worked alongside other pre-existing signals rather than replacing them. While participants acknowledged that citations themselves could harbor biases, they still considered them invaluable for fact-checking and validation. Intriguingly, some participants proposed the incorporation of a user-rating metric for individual citations, reflecting an eagerness to refine the evaluation process further. As noted in Section \ref{background}, a group of users collectively liking or sharing a news article on social media signals collective trust in the information's credibility \cite{hong2006influence,flanagin2008digital}. Provided this finding proves true in Viblio's application, the additional user-rating metric could provide another signal that users could employ in their credibility-determination process. This revelation underscores Viblio's potential as a catalyst for greater user engagement and discernment in the face of misinformation.

Despite Viblio's potential as a tool against misinformation, it is still susceptible to issues other crowd-based or social platforms face. In the current version of Viblio, anyone, including bad actors, can add citations. In some cases, this could lead to a video, full of misinformation, citing a multitude of untrustworthy and/or incorrect citations. To combat this, it is critical to safeguard against low-quality and malicious citations. While this study focused solely on implementing an initial system for trust and credibility, it is imperative to integrate moderation tools in the future. Similar studies and projects, like X's Community Notes, can be used as an example to develop Viblio further. 

One possible approach that Viblio can follow is that of Wikipedia's crowd-sourced citations, which set standards contributors must follow \cite{Wikipedia_2023}. Another system to look to is Community Notes, which has already implemented both a reputation system for users and a voting system for individual 'notes'. So far, Community Notes has been shown to increase the quality of content produced and decrease the number of people sharing content \cite{borwankar_democratization_2022}. Preliminary findings from Community Notes suggest a similar success for Viblio if implemented at scale. It is important to note the differences that Viblio faces in comparison to similar tools on other platforms. While Community Notes focuses on small amounts of text-based data, Viblio must be able to address multiple viewpoints and information points for each video. One tweet may only touch on one point, but a YouTube video has the possibility to address a range of opinions and facts. Viblio has the ability to provide viewers with a way to make an overarching opinion on the credibility of the video and to determine the validity of individual statements. Viblio also faces challenges when compared to Community Notes or Wikipedia in terms of the video format; citations for text have long-established norms and formats.

Another challenge area to consider for Viblio is in situations where the facts are not settled yet, or information is contested. In these situations, citations claiming to support or refute a fact may only confuse the sense-making process of users. This struggle has already become apparent on Community Notes \cite{Fan_Dottle_Wagner_2022}. One way Community Notes is attempting to combat this is by showing the community multiple notes suggested by users as well as upvotes to allow for the community to collectively decide what to eventually share. Given our participant's use of like/dislike counters as a credibility signal in the explorative study already, a possible approach for Viblio is to include individual counters for each citation on a video.

\subsection{Scalability}\label{scalability}
A question that remains in regard to Viblio is the possible scalability of the application. The crowd-sourced approach that Viblio currently implements requires a high degree of user involvement in comparison to systems like YouTube's information panels. However, unlike information panels, Viblio is able to provide customized approaches to each individual video that can mitigate issues found with information panels (see Section 3.2.1). Viblio, like Community Notes, could prove its strength for high-traffic videos, where more users are likely to add citations throughout a video. On the other hand, Viblio's current system falls short for low-traffic videos which will be unlikely to have many (or any) citations added. 

This shortcoming could be mitigated through the possibility of automating citations. While AI-generated citations would help combat low-traffic misinformation, the citations themselves may prove to be another possible source of misleading sources or misinformation. An option that warrants further exploration is a human-AI collaborative approach, where AI suggests possible citations that users approve or deny. Another future possibility AI can add is the match of current citations to other matches or near matches of claims/video sequences across different videos, allowing for a citation to be copied over.

\subsection{Balancing Cognitive Load and Credibility}

An unexpected but enlightening finding from our study pertains to the potential mental load Viblio introduces into the user experience. Some participants noted that using Viblio alongside video content required a noticeable additional cognitive effort compared to their customary YouTube viewing habits. However, the unanimous consensus among all participants was that this added cognitive load was justified, particularly in the context of complex and controversial topics \cite{jahanbakhsh_exploring_2021, pennycook_lazy_2019}. This observation highlights Viblio's role as a tool for enhancing critical thinking and information literacy, potentially fostering more thoughtful engagement with digital content.

Beyond its primary goal of aiding credibility determination, our study revealed that Viblio can serve a broader set of purposes, particularly in the realm of education. Participants envisioned Viblio as a valuable supplement to video content, allowing viewers to access in-depth sources and expand their knowledge on specific topics. For instance, in educational contexts such as computer science history, where videos often provide concise overviews, Viblio's ability to offer quick and easy access to comprehensive sources can significantly enhance the learning experience. As educational content continues to proliferate on platforms like YouTube, Viblio stands as a promising tool with the potential to elevate the quality and depth of learning outcomes.

\subsection{Design Considerations and User Feedback}

Our findings also shed light on various design choices and enhancements that could further refine Viblio's utility. Participants' feedback suggests the importance of exploring different placements for citations, adding filter views to facilitate the review of citations by type, and implementing a minimum show-time for citations, even if the cited period is shorter. Moreover, the idea of crowdsourcing the usefulness of citations, as proposed by one participant, holds promise in augmenting Viblio's functionality, especially given preliminary findings for Community Notes \cite{borwankar_democratization_2022}. Users' varying responses to citations based on their initial credibility ratings emphasize the need for Viblio to cater to different user needs and perceptions.

\section{Future Work}\label{future-work}
Looking forward, the work on Viblio has many possible directions to grow in. As mentioned in Section \ref{scalability}, the scalability of this system is unknown. While systems like Community Notes can be looked at as an example, there is no way to know the greater impact of Viblio until it is tested on a greater scale. Also, given participants' mixed views on Viblio’s use in political environments (Section \ref{useascred}), a broader study on the use of citations by political groups on content both aligning and disagreeing with their views could provide crucial knowledge for combating political misinformation. Another major area for work, mentioned in Section \ref{viblioDesign}, is to evaluate how to guard against low-quality and/or malicious contributions in a crowd-sourced system such as Viblio. Again, similar systems such as Community Notes provide an example of how Viblio could address this in the future.

Viblio also has the opportunity to expand into the search results page on YouTube, providing crucial credibility-related information to viewers when deciding which video to watch. Our exploratory study (Section \ref{exploratory-study}) explored how the ranking of search results, as well as other credibility signals, could influence viewer's choices. Future work can explore how new design implementations of Viblio specifically for the search results page could possibly influence viewer's choices. If Viblio expands into a reputable source, the possibility of citations influencing the ranking, recommendation, moderation, or monetization algorithms/processes becomes available. A trade-off of this may be an increase in the creation of false citations or the motivation of bad actors to game the system. Previous work has also shown that stance-based labels may intensify selective exposure and may make users more vulnerable to polarised opinions and fake news, so careful consideration will have to be taken in the design process \cite{10.1145/3274324}.

Another direction to explore is the integration of Viblio on different social platforms. During our studies, multiple participants commented on the utility of Viblio on other platforms; for example: ``\textit{there's a Korean social website called Naver and all Koreans use that and add to that and stuff. So if I had that on Naver, it will be, like, very, very helpful}'' (Reflective Study P8). Another key player in the spread of information online is now TikTok, as there is a growing demographic of users turning to TikTok for their news \cite{matsa_more_nodate}. On TikTok, there are even fewer recognizably trustworthy sources producing content. While the current design of Viblio would not be able to directly translate, the application of a video citation system on TikTok is a necessary next step.

\section{Conclusion}\label{conclusion}
Viblio's introduction as a prototype system for YouTube has unlocked new dimensions in user-driven credibility determination. While we achieved our initial goal of empowering users to assess the credibility of video content, our reflective study also illuminates Viblio's potential for multifaceted applications, ranging from educational enhancement to the fostering of critical thinking. Viblio goes beyond merely providing citations and offers a crucial evaluative point, facilitating fact-checking and validation. This user-driven approach has the potential to enhance user engagement and discernment, which are vital tools in combating misinformation. As we continue to refine and expand Viblio, these findings will guide its development, ensuring that it remains a valuable asset in the ever-evolving landscape of digital information and online video content evaluation.

\begin{acks}
This work is supported by the WikiCred Grants Initiative. Thank you to Anson Huang for his dedicated work on this project.
\end{acks}

\bibliographystyle{ACM-Reference-Format}
\bibliography{biblio}

\appendix
\begin{table*}[]
\caption{Participant Interaction Counts by Video and Category}\label{appTable1}
\begin{tabular}{c|c|cc|c}
\hline
\multirow{3}{*}{Video Titles by Category}                                                                                                       & \multirow{3}{*}{\begin{tabular}[c]{@{}c@{}}Number of \\ Participant \\ Responses\end{tabular}} & \multicolumn{2}{c|}{Interactions with Interface}                                 & \multirow{3}{*}{\begin{tabular}[c]{@{}c@{}}Click-Through \\ to Citation \\ Article\end{tabular}} \\ \cline{3-4}
                                                                                                                                                &                                                                                                & \multicolumn{1}{c|}{\multirow{2}{*}{Timeline View}} & \multirow{2}{*}{List View} &                                                                                                  \\
                                                                                                                                                &                                                                                                & \multicolumn{1}{c|}{}                               &                            &                                                                                                  \\ \hline
\multirow{2}{*}{Educational (avg)}                                                                                                              & \multirow{2}{*}{9.57}                                                                          & \multicolumn{1}{c|}{\multirow{2}{*}{25.43}}         & \multirow{2}{*}{14.71}     & \multirow{2}{*}{3.43}                                                                            \\
                                                                                                                                                &                                                                                                & \multicolumn{1}{c|}{}                               &                            &                                                                                                  \\ \hline
Cereal - A Brief History - The New Yorker                                                                                                       & 5                                                                                              & \multicolumn{1}{c|}{13}                             & 0                          & 0                                                                                                \\
\begin{tabular}[c]{@{}c@{}}Comparative Anatomy: What Makes Us Animals \\ - Crash Course Biology \#21\end{tabular}                               & 11                                                                                             & \multicolumn{1}{c|}{23}                             & 12                         & 6                                                                                                \\
Heredity: Crash Course Biology \#9                                                                                                              & 11                                                                                             & \multicolumn{1}{c|}{41}                             & 3                          & 6                                                                                                \\
How does laser eye surgery work? - Dan Reinstein                                                                                                & 12                                                                                             & \multicolumn{1}{c|}{46}                             & 34                         & 3                                                                                                \\
Is Organic Food Worse For You?                                                                                                                  & 5                                                                                              & \multicolumn{1}{c|}{7}                              & 0                          & 2                                                                                                \\
mRNA vaccines, explained                                                                                                                        & 12                                                                                             & \multicolumn{1}{c|}{28}                             & 38                         & 5                                                                                                \\
When a physics teacher knows his stuff !!                                                                                                       & 11                                                                                             & \multicolumn{1}{c|}{20}                             & 16                         & 2                                                                                                \\ \hline
\multirow{2}{*}{News (avg)}                                                                                                                     & \multirow{2}{*}{11.38}                                                                         & \multicolumn{1}{c|}{\multirow{2}{*}{20.13}}         & \multirow{2}{*}{6.75}      & \multirow{2}{*}{2.63}                                                                            \\
                                                                                                                                                &                                                                                                & \multicolumn{1}{c|}{}                               &                            &                                                                                                  \\ \hline
Biden says he's 'planning on running in 2024'                                                                                                   & 10                                                                                             & \multicolumn{1}{c|}{16}                             & 7                          & 2                                                                                                \\
Biden: ‘I plan on running’ in 2024 presidential election                                                                                        & 12                                                                                             & \multicolumn{1}{c|}{14}                             & 2                          & 3                                                                                                \\
California investigates ExxonMobil over its plastics waste                                                                                      & 12                                                                                             & \multicolumn{1}{c|}{25}                             & 20                         & 2                                                                                                \\
Fake heiress Anna Sorokin to be deported l GMA                                                                                                  & 12                                                                                             & \multicolumn{1}{c|}{35}                             & 2                          & 2                                                                                                \\
Fox settles Dominion case, but bigger lawsuit looms                                                                                             & 11                                                                                             & \multicolumn{1}{c|}{23}                             & 6                          & 2                                                                                                \\
\begin{tabular}[c]{@{}c@{}}Kentucky Gov. Beshear lost 'close friend' in \\ Louisville shooting\end{tabular}                                     & 12                                                                                             & \multicolumn{1}{c|}{23}                             & 11                         & 7                                                                                                \\
\begin{tabular}[c]{@{}c@{}}Picket lines form at Rutgers University campuses \\ during historic strike\end{tabular}                              & 10                                                                                             & \multicolumn{1}{c|}{6}                              & 2                          & 0                                                                                                \\
\begin{tabular}[c]{@{}c@{}}US secrets leaked in social media post containing \\ Ukraine, Russia documents\end{tabular}                          & 12                                                                                             & \multicolumn{1}{c|}{19}                             & 4                          & 3                                                                                                \\ \hline
\multirow{2}{*}{Controversial (avg)}                                                                                                            & \multirow{2}{*}{9.70}                                                                          & \multicolumn{1}{c|}{\multirow{2}{*}{17.60}}         & \multirow{2}{*}{5.50}      & \multirow{2}{*}{3.90}                                                                            \\
                                                                                                                                                &                                                                                                & \multicolumn{1}{c|}{}                               &                            &                                                                                                  \\ \hline
\begin{tabular}[c]{@{}c@{}}"COUNT ONLY LEGAL VOTES" Rudy Giuliani OPENING \\ STATEMENT On Election Fraud Claims | NewsNOW From FOX\end{tabular} & 9                                                                                              & \multicolumn{1}{c|}{13}                             & 6                          & 3                                                                                                \\
Deadline on abortion pill ruling                                                                                                                & 10                                                                                             & \multicolumn{1}{c|}{7}                              & 0                          & 2                                                                                                \\
GOP ban on trans student athletes passes House                                                                                                  & 10                                                                                             & \multicolumn{1}{c|}{18}                             & 0                          & 5                                                                                                \\
\begin{tabular}[c]{@{}c@{}}Ilhan Omar connected Ballot Harvester in cash-for-ballots \\ scheme: "Car is full" of absentee ballots\end{tabular}  & 8                                                                                              & \multicolumn{1}{c|}{7}                              & 4                          & 5                                                                                                \\
Joe Biden’s America shouldn’t be our future: Matthew Whitaker                                                                                   & 11                                                                                             & \multicolumn{1}{c|}{23}                             & 7                          & 4                                                                                                \\
\begin{tabular}[c]{@{}c@{}}Leak of classified documents and intelligence \\ shakes U.S. Department of Defense\end{tabular}                      & 11                                                                                             & \multicolumn{1}{c|}{21}                             & 0                          & 7                                                                                                \\
\begin{tabular}[c]{@{}c@{}}Major anti-abortion group gives scathing response \\ to Trump comments\end{tabular}                                  & 8                                                                                              & \multicolumn{1}{c|}{39}                             & 8                          & 4                                                                                                \\
\begin{tabular}[c]{@{}c@{}}Ruling Against Abortion Pill Mifepristone Could \\ Fundamentally Alter The FDA, Becerra Warns\end{tabular}           & 12                                                                                             & \multicolumn{1}{c|}{34}                             & 26                         & 5                                                                                                \\
\begin{tabular}[c]{@{}c@{}}WH 'disagrees strenuously' with Texas judge's \\ decision on abortion pill\end{tabular}                              & 10                                                                                             & \multicolumn{1}{c|}{4}                              & 3                          & 3                                                                                                \\
What Did Bill Gates Say About COVID Vaccine Side Effects?                                                                                       & 8                                                                                              & \multicolumn{1}{c|}{10}                             & 1                          & 1                                                                                                \\ \hline
\end{tabular}
\end{table*}

\begin{table*}[]
\caption{Participant Interaction Counts by Participant}\label{appTable2}
\begin{tabular}{c|cc|c|c}
\hline
\multirow{2}{*}{Participant} & \multicolumn{2}{c|}{Interactions with Interface} & \multirow{2}{*}{\begin{tabular}[c]{@{}c@{}}Click-through to \\ citation article\end{tabular}} & \multirow{2}{*}{Total} \\ \cline{2-3}
                             & \multicolumn{1}{c|}{List View}  & Timeline View  &                                                                                               &                        \\ \hline
1                            & \multicolumn{1}{c|}{0}              & 74         & 2                                                                                             & 76                     \\
2                            & \multicolumn{1}{c|}{0}              & 49         & 1                                                                                             & 50                     \\
3                            & \multicolumn{1}{c|}{6}              & 0          & 0                                                                                             & 6                      \\
4                            & \multicolumn{1}{c|}{0}              & 6          & 0                                                                                             & 6                      \\
5                            & \multicolumn{1}{c|}{14}             & 19         & 1                                                                                             & 34                     \\
6                            & \multicolumn{1}{c|}{85}             & 6          & 5                                                                                             & 96                     \\
7                            & \multicolumn{1}{c|}{0}              & 73         & 4                                                                                             & 77                     \\
8                            & \multicolumn{1}{c|}{0}              & 18         & 10                                                                                            & 28                     \\
9                            & \multicolumn{1}{c|}{0}              & 11         & 3                                                                                             & 14                     \\
10                           & \multicolumn{1}{c|}{3}              & 126        & 35                                                                                            & 164                    \\
11                           & \multicolumn{1}{c|}{69}             & 7          & 1                                                                                             & 77                     \\
12                           & \multicolumn{1}{c|}{0}              & 101        & 20                                                                                            & 121                    \\ \hline
Total                        & \multicolumn{1}{c|}{212}            & 515        & 84                                                                                            & 811                    \\ \hline
\end{tabular}
\end{table*}

\end{document}